\documentclass[10pt]{article}  
\usepackage{graphicx}
\usepackage{amssymb}
\input epsf
\parskip=\medskipamount
\def\al{\alpha}
\def\be{\beta}
   \def\Ga{\Gamma}

\def\la{\lambda}  
\def\k{\kappa}  
\def\kp{\kappa}  
   
\def\om{\omega}   
      
\def\IB{\relax{\rm l\kern-.18 em B}}
\def\IC{\relax{\rm l\kern-.50 em C}}
\def\IE{\relax{\rm l\kern-.12 em E}}
\def\IH{\relax{\rm l\kern-.18 em H}}
\def\IK{\relax{\rm l\kern-.18 em K}}
\def\IL{\relax{\rm I\kern-.18 em L}}
\def\IN{\relax{\rm I\kern-.18 em N}}
\def\IR{\relax{\rm I\kern-.18 em R}}

\def\smallonehalf{\frac{{}_1}{{}^2}}
 
\def\\{\hfill\break}
\def\smallonehalf{\frac{{}_1}{{}^2}}
 
\def\wt{\widetilde}

\def\k{\kappa}                       
\font\tenfrak=eufm10  \font\sevenfrak=eufm7  \font\fivefrak=eufm5
\newfam\frakfam
\textfont\frakfam=\tenfrak
\scriptfont\frakfam=\sevenfrak
\scriptscriptfont\frakfam=\fivefrak

\newtheorem{proposicion}{Proposition}

\def\wt{\widetilde}
\def\frac#1#2{{#1\over #2}}
\def\fracpd#1#2{\frac{\partial #1}{\partial #2}}
\def\ptos{\leaders\hbox to 2mm{\hfil{.}\hfil}\hfill}

\begin{document}

\title{ Superintegrable deformations of superintegrable systems :    \\ [2pt]
Quadratic superintegrability and higher-order superintegrability}

\author{ Manuel F. Ra\~nada$^{\,a)}$  \\ [3pt]
 {\sl Dep. de F\'{\i}sica Te\'orica and IUMA } \\
 {\sl Universidad de Zaragoza, 50009 Zaragoza, Spain}     }

\date{}
\maketitle 

{\bigskip}

\begin{abstract} 
The superintegrability of four Hamiltonians $\wt{H_r} = \la\, H_r$, $r=a,b,c,d$, where  $H_r$ are known Hamiltonians and $\la$ is a certain function defined on the configuration space and depending of a parameter $\kp$, is studied. 
The new Hamiltonians, and the associated constants of motion $J_{ri}$, $i=1,2,3$, are continous functions of the parameter $\kp$. 
The first part is concerned with separability and quadratic superintegrability (the integrals of motion are quadratic in the momenta) and the second part is devoted to the existence of higher-order superintegrability. The results obtained in the second part are related with the TTW and the PW systems. 

\end{abstract}

\begin{quote}
Higher-order  constants of motion. Complex factorization.  Integrability on curved spaces.  

{\smallskip}

{\sl Running title:}{\enskip}
New families of superintegrable systems.

{\smallskip}
AMS classification:  37J35 ; 70H06
{\smallskip}

\end{quote}

{\vfill}
\vfill
\footnoterule{\small
\begin{quote}
$^{a)}$  {\tt E-mail: {mfran@unizar.es}  }
\end{quote}
}

\newpage


\section{Introduction }

A Hamiltonian system with $n$ degrees of freedom is called Liouville (or Liouville-Arnold) integrable if it is endowed with $n$ functionally independent integrals of motion in involution (including the Hamiltonian itself). Some integrable systems, as the harmonic oscillator or the Kepler problem,  admit more constants of motion than degrees of freedom; they are called superintegrable.  
Therefore a Hamiltonian with two degrees of freedom is said to be superintegrable if it admits two fundamentals integrals of motion, $I_1$ and $I_2$, that Poisson commute and a third independent integral $I_3$.  The additional integral $I_3$ has vanishing Poisson bracket with $H$ but not necessarily with $I_1$ and $I_2$. 

It is known that if a system is separable (Hamilton-Jacobi separable in the classical case or Schr\"odinger separable in the quantum case) then it  is integrable with integrals of motion of at most second order in momenta.   
Thus, if a system admits multiseparability (separability in several different systems of coordinates) then it is endowed with ``quadratic superintegrability"  (superintegrability with linear or quadratic integrals of motion). 

 Fris et al. studied \cite{FrMaSmUW65}  the two-dimensional Euclidean systems, which admit  separability in more than one coordinate systems and obtained four families of potentials $V_r$, $r = a, b, c, d$, possessing three functionally independent integrals of motion  (they were mainly interested in the quantum two-dimensional Schr\"odinger equation but the results obtained are also valid at the classical level). Then other authors studied similar problems on higher-dimensional Euclidean spaces \cite{Ev90Pra}--\cite{KaWiMiPo99}, on two-dimensional spaces with a pseuo-Euclidean metric (Drach potentials) \cite{Ra97Jmp}--\cite{Camp14}, or on curved spaces \cite{GrPoSi95b}--\cite{GonKas14AnnPhys} (see \cite{MiPWJPa13} for a recent review on superintegrability that includes a long list of references).

 For some time the studies on superintegrability were mainly concerned  with ``quadratic superintegrability" but recent studies have proved the existence of certain systems endowed with Ôhigher order superintegrabilityÕ, that is, with integrals of motion which are polynomials in the momenta of higher order than two. 
We mention the Calogero-Moser system whose superintegrability is related with a Lax equation \cite{Woj83}--\cite{Ra99Jmp}  (this formalism is not considered in this paper) and three important systems that are separable but in only one system of coordinates, the generalized SW system system (caged anisotropic oscillator) \cite{EvVe08b}--\cite{RaRoSa10}, the Tremblay-Turbiner-Winternitz (TTW) system \cite{TTW09}--\cite{Ra14JPaTTWk}, and the Post--Winternitz (PW) system \cite{PostWint10}--\cite{Ra13JPaPW}; in these three cases two of the integrals are quadratic but the third one is of higher order.

One important point is that although the number of superintegrable systems can be considered as rather limited, they are not  however isolated systems but, on the contrary, they frequently appear grouped in families; for example, everyone of the above mentioned potentials $V_r$, $r=a,b,c,d$, has structure of a  three-dimensional vector space. 
Now in this paper we consider the idea of  one-parameter deformations of a given Hamiltonian, that is,  families of Hamiltonians depending of a real parameter $\kp$ that are superintegrable for all the values of $\kp$ (in the domain of the parameter) and that for $\kp=0$ they reduce to superintegrable Hamiltonians previously studied. 

The main objective of this article is twofold. First, study the existence of superintegrable deformations of the four two-dimensional Euclidean systems (some of the results obtained are related with some nonlinear systems studied in  \cite{BurgosPhysd08}--\cite{BurgosSigma11}) and, second,  study the existence of superintegrable deformations of the TTW and the PW systems. 
In these two cases we prove the superintegrability and we obtain the explicit expression of the third integral (that we recall is of higher-order) as the product of  powers  of two particular complex functions (this complex formalism is very similar to the approach presented in  \cite{Ra12JPaTTW},  \cite{Ra14JPaTTWk}, for the TTW system and in 
\cite{Ra13JPaPW}  for the PW system)

  The plan of the article is as follows:
Sec. 2 is devoted to recall the main characteristics of the four two-dimensional potentials whith separability in two different coordinate systems in the Euclidean plane. Then Sec. 3 and 4  are concerned with quadratic superintegrability and Sec. 5 with higher order constants of motion.
In Sec. 3 we first introduce the idea of deformation (depending of one parameter $\kp$) of a Hamiltonian and then we study four families of  superintegrable families endowed with multiple separability. 
Sec. 5 has two parts. 
In the first part we study a $\kp$-dependent Hamiltonian related with the harmonic oscillator,  that can be considered as a deformation of the TTW system, and in the second part a $\kp$-dependent Hamiltonian related with the Kepler problem, that can be considered as a deformation of the PW system. 
Finally in Sec. 6 we make some final comments.

\section{Superintegrability with quadratic constants of motion in the Euclidean plane }

Let us denote by   $V_r$, $r=a,b,c,d$,  the four two-dimensional potentials whith separability in two different coordinate systems in the Euclidean plane.

   The two first potentials, $V_a$ y $V_b$, are related with the harmonic oscillator. They satisfy the equation $V_{xy}=0$ and correspond, therefore, to a direct sum of one-degree of freedom systems.

\begin{enumerate}
\item[(a)]  The following potential  
\begin{equation}   
  V_a =   ({\smallonehalf})\al^2 (x^2 + y^2) + \frac{k_2}{x^2}  + \frac{k_3}{y^2}
\end{equation}
  is separable in  (i) Cartesian coordinates and  (ii) polar coordinates. 
The three constants of motion, $I_{a1}$, $I_{a2}$, y $I_{a3}$, are given by    
\begin{eqnarray*}
  I_{a1} &=&  p_x^2  + \al^2 x^2 + \frac{2k_2}{x^2}\,, \cr
  I_{a2} &=&  p_y^2  +  \al^2 y^2 + \frac{2k_3}{y^2}\,, \cr
  I_{a3} &=&  (x p_y - y p_x)^2  +  2k_2\bigl(\frac{y}{x}\bigr)^2 +
  2k_3\bigl(\frac{x}{y}\bigr)^2\,. 
\end{eqnarray*}

\item[(b)]  The following potential  
\begin{equation}  
  V_b = ({\smallonehalf})\al^2 (4 x^2 + y^2) + \frac{k_2}{x^2}  + k_3 x
\end{equation}
is separable in  (i) Cartesian coordinates and  (ii) parabolic coordinates. 
The three constants of motion,  $I_{b1}$, $I_{b2}$, y $I_{b3}$, are given by \begin{eqnarray*}
  I_{b1} &=& p_x^2 +  4 \al^2 x^2 + 2 k_3 x\,, \cr
  I_{b2} &=&  p_y^2 +  \al^2 y^2 + \frac{2 k_2}{y^2} \,, \cr
  I_{b3} &=&  (x p_y - y p_x) p_y  -  \al^2 x y^2 + k_2 \bigl(\frac{2 x}{y^2}\bigr) - \frac{k_3}{2y^2}
  \,. 
\end{eqnarray*}

\end{enumerate}

  The two other potentials, $V_c$ y $V_c$, are related with the Kepler problem. 

\begin{enumerate}
\item[(c)]  The following potential  
\begin{equation}   
  V_c =  \frac{k_1}{\sqrt{x^2 + y^2}}  +  \frac{k_2}{y^2}  +  \frac{k_3x}{y^2 \sqrt{x^2 + y^2}}
\end{equation}   
is separable in  (i) polar coordinates and  (ii) parabolic coordinates. 
The first constant of motion is the Hamiltonian itself, that is $ I_{c1} =  H_c$, and the other two constants of motion,  $I_{c2}$, and $I_{c3}$, are given by   
\begin{eqnarray*}
  I_{c2} &=&  (x p_y - y p_x)^2 + \frac{2 k_2 x^2}{y^2} +
  \frac{2 k_3 x \sqrt{x^2 + y^2}}{y^2}  \,, \cr
  I_{c3}&=& (x p_y - y p_x) p_y + \frac{k_1 x}{\sqrt{x^2 + y^2}}
  + \frac{2 k_2 x}{y^2} + \frac{k_3 (2 x^2 + y^2)}{y^2 \sqrt{x^2 +
y^2}}   \,.  
\end{eqnarray*}

\item[(d)]  The following potential  
\begin{equation}   
V_d =  \frac{k_1}{\sqrt{x^2 + y^2}}   
  + k_2  \frac{ \bigl[\sqrt{x^2 + y^2} + x\bigr]^{1/2} }{ \sqrt{x^2 + y^2}}   
  + k_3 \frac{ \bigl[\sqrt{x^2 + y^2} - x\bigr]^{1/2} }{ \sqrt{x^2 + y^2}}
\end{equation}   
 is separable in  (i) parabolic coordinates $(a,b)$ and  (ii) a second system of parabolic coordinares $(\alpha,\beta)$ obtained from $(a,b)$ by a rotation. 
The first constant of motion is the Hamiltonian itself, that is $ I_{d1} =  H_d$, and the other two constants of motion,  $I_{d2}$, and $I_{d3}$, are given by  
 \begin{eqnarray*}
  I_{d2}  &=&   (x p_y - y p_x) p_y + \frac{k_1 x}{\sqrt{x^2 + y^2}}
- k_2  \frac{y\, \bigl[\sqrt{x^2 + y^2} - x\bigr]^{1/2} }{\sqrt{x^2 + y^2}} 
+ k_3 \frac{y\, \bigl[\sqrt{x^2 + y^2} + x\bigr]^{1/2} }{\sqrt{x^2 + y^2}}  \,,     \cr
  I_{d3}  &=&  (x p_y - y p_x) p_x - \frac{k_1 y}{\sqrt{x^2 + y^2}}
- k_2  \frac{x\, \bigl[\sqrt{x^2 + y^2} - x\bigr]^{1/2} } {\sqrt{x^2 + y^2}}  
+ k_3 \frac{x\, \bigl[\sqrt{x^2 + y^2} + x\bigr]^{1/2} } {\sqrt{x^2 + y^2}} \,.     
\end{eqnarray*}
\end{enumerate}

\section{Four new superintegrable families endowed with multiple separability  }

Suppose we are given a  Hamiltonian $H$; then we can construct a new Hamiltonian  $ \wt{H}$ as $\wt{H} = \la\, H$ where $\la$ is a certain function defined on the configuration space. This new Hamiltonian represents a new and different  dynamics; for example if $H$ is defined on an Euclidean space then the new dynamics will be nonEuclidean.  More particularly, we are interested in multipliers $\lambda$ that preserve certain properties of $H$ as integrability or separability. 
For example if $H$ is defined in the Euclidean plane and is separable in Cartesian coordinates $(x,y)$ and $\la$ is of the form $\la=1/\mu$, $\mu = f(x) + g(y)$, then $\wt{H}$ is separable in Cartesian coordinates as well, and if $H$ is separable in polar coordinates $(r,\phi)$ and $\la$ is of the form $\la=1/\mu$, $\mu = f(r) + g(\phi)/r^2$, then $\wt{H}$ is also separable in polar coordinates. A more strong condition is that $\lambda$ must preserve not just separability but multiple separability; this requirement will strongly restrict the form of the multiplier. 

Another important property we wish to introduce is that the new Hamiltonian $ \wt{H}$ must be a deformation of $H$.   By deformation we mean that $\lambda$, and therefore $\wt{H}$, will depend of a parameter $\kp$ in such a way that 
\begin{itemize}
\item[(i)]   The new Hamiltonian $\wt{H}(\kp)$  must be a continous function of $\kp$ (in a certain domain of the parameter). 
\item[(ii)]  When $\kp\to 0$ we have $\lambda\to 1$ and then the dynamics of the Euclidean Hamiltonian $H$ is recovered.
\end{itemize}

In this section we study the separability and the superintegrability of four Hamiltonians $\wt{H_r}(\kp)$ obtained as deformations of the four Hamiltonians $H_r$, $r=a,b,c,d$.

\subsection{Hamiltonian $\wt{H_a}(\kp)$ }

Let us consider the Hamiltonian  $H_a$ 
\begin{equation}
  H_a = \frac{1}{2}\,\bigl(p_x^2 + p_y^2\bigr)  + \frac{1}{2}\,\al^2\,\bigl(x^2 + y^2\bigr)  +  \frac{k_2}{x^2}   +  \frac{k_3}{y^2} 
\end{equation}
and denote by  $\la_a$ the following multiplier
$$
 \la_a  = \frac{1}{\mu_a}\,,{\quad}   \mu_a(x,y) = 1 - \kp\,r^2 \,,{\quad} 
 r^2 = x^2+y^2 \,. 
$$
Then the new $\kp$-dependent Hamiltonian $\wt{H_a}(\kp)$ defined as $\wt{H_a}(\kp) = \la_a H_a$ is given by 
\begin{equation}
 \wt{H_a}(\kp) = \frac{1}{2}\,\Bigl(\frac{p_x^2 + p_y^2}{1 - \kp\,r^2}\Bigr)  + \frac{1}{2}\,\al^2\,\Bigl(\frac{x^2 + y^2}{1 - \kp\,r^2}\Bigr)  +  \frac{k_2}{(1 - \kp\,r^2)\,x^2}   +  \frac{k_3}{(1 - \kp\,r^2)\,y^2}   \,.
\end{equation}

The parameter $\kappa$ can take both positive and negative values. In the $\kappa<0$ case the $\wt{H_a}(\kp)$ dynamics is correctly defined for all the values of the variables; 
nevertheless when $\kappa>0$, the Hamiltonian (and the associated  dynamics)
has a singularity at $1 -\,\kp\,r^2=0$, so in this case the  $\wt{H_a}(\kp)$
dynamics is defined in the interior of the circle $r^2=1/\kp$, $\kp>0$,  that is the region in which kinetic term is positive definite.

\begin{itemize}

\item[(i)] {Cartesian separability }

The Hamilton-Jacobi (H-J) equation takes the form 
$$
\frac{1}{2 }\,\Bigl(\frac{1}{1 - \k\,r^2}\Bigr) \Bigl[\Bigl(\fracpd{W}{x}\Bigr)^2 + \Bigl(\fracpd{W}{y}\Bigr)^2\Bigr]  + \frac{1}{2}\,\al^2\,\Bigl(\frac{x^2 + y^2}{1 - \kp\,r^2}\Bigr) +  \frac{k_2}{(1 - \kp\,r^2)\,x^2}  +  \frac{k_3}{(1 - \kp\,r^2)\,y^2} = E
$$
so that if we assume that $W$ can be written as $W=W_x(x)+W_y(y)$ then we can perform a separation of variables and arrive to the following two one-variable expressions 
\begin{eqnarray*}
 \Bigl[\Bigl(\fracpd{W_x}{x}\Bigr)^2 +  \al^2 x^2 +  \frac{2 k_2}{x^2} + 2 \kp E x^2\Bigr]   &=&  K +  E  \,, \cr 
\Bigl[\Bigl(\fracpd{W_y}{y}\Bigr)^2 +  \al^2 y^2 +  \frac{2 k_3}{y^2} + 2 \kp E y^2\Bigr] &=&  -\,K +  E   \,,
\end{eqnarray*}
where $K$ denotes the constant associated to the separability. Everyone of these two expressions determine a constant of motion. So  the $\kp$-dependent  functions $J_{a1}$ and $J_{a2}$ given by 
\begin{eqnarray*}
 J_{a1}  &=& p_x^2 +  \al^2 x^2 +  \frac{2 k_2}{x^2} + 2  \kp \wt{H_a} x^2  \,,\cr 
 J_{a2}  &=& p_y^2 + \al^2 y^2 +  \frac{2 k_3}{y^2} + 2  \kp \wt{H_a} y^2   \,,  
\end{eqnarray*}
are constants of motion satisfying the following properties 
$$
 (i)\   dJ_{a1}\,\wedge\, dJ_{a2}\ne 0 \,,{\quad} 
 (ii)\   \{J_{a1}\,,\, J_{a2}\} = 0 \,,{\quad} 
 (iii)\   \wt{H_a}(\kp) = \frac{1}{2}\bigl(J_{a1} + J_{a2} \bigr) \,. 
$$

\item[(ii)] {Polar separability }

The Hamilton-Jacobi (H-J) equation takes the form 
$$
\frac{1}{2m}\,\Bigl[\Bigl(\fracpd{W}{r}\Bigr)^2 + \frac{1}{r^2}\Bigl(\fracpd{W}{\phi}\Bigr)^2\Bigr]  +  \al^2\,r^2  +  \frac{k_2}{r^2\cos^2\phi}  +  \frac{k_3}{r^2\sin^2\phi}  = (1 - \kp\,r^2) E
$$
so that if we assume that $W$ is of the form $W=W_r(r)+W_\phi(\phi)$ we can perform a separation of variables and rewrite the equation as the sum of two one-variable summands
$$
 \Bigl[ r^2\Bigl(\fracpd{W_r}{r}\Bigr)^2 + \al^2 r^4   - 2  E r^2(1 - \k\,r^2) \Bigr]+    
 \Bigl[  \Bigl(\fracpd{W_\phi}{\phi}\Bigr)^2 +  \frac{2  k_2}{\cos^2\phi}  +  \frac{2  k_3}{\sin^2\phi} \Bigr] = 0    
$$
so that the following function
$$
 J_{a3} = p_\phi^2 +  \frac{2  k_2}{\cos^2\phi}  +  \frac{2  k_3}{\sin^2\phi} 
$$
is a constant of motion. 

\end{itemize}

We summarize the results in the following proposition. 

\begin{proposicion} 
The $\kp$-dependent Hamiltonian  $\wt{H_a}(\kp)$ is H-J separable in Cartesian $(x,y)$ and polar $(r,\phi)$ coordinates  and  it is endowed with the following  three quadratic constants of motion 
\begin{eqnarray*}
 J_{a1}  &=& \frac{1}{(1 - \kp\,r^2)} \Bigl( (1 - \kp y^2) p_x^2 + \k x^2 p_y^2 +  \al^2 x^2   + \frac{2 k_2(1 - \kp y^2)}{x^2} + \frac{2 k_3x^2}{y^2}\Bigr)   \,, \cr
 J_{a2} &=& \frac{1}{(1 - \kp\,r^2)} \Bigl( (1 - \kp x^2) p_y^2 + \k y^2 p_x^2 +  \al^2 y^2 + \frac{2 k_2y^2}{x^2} + \frac{2 k_3(1 - \kp x^2)}{y^2} \Bigr) \,, \cr
 J_{a3} &=&  (x p_y - y p_x)^2  +  2 k_2\bigl(\frac{y}{x}\bigr)^2 +  2 k_3\bigl(\frac{x}{y}\bigr)^2\,.  
\end{eqnarray*}
\end{proposicion}

We note that the two first functions satisfy the correct limit when $\kp\to 0$, that is   $\lim_{\kp\to 0} \,J_{ai} = I_{ai}$,  $i=1,2$. 
Concerning the third function $J_{a3}$ it is $\kp$-independent and it coincides with the original constant  $I_{a3}$.

\subsection{Hamiltonian  $\wt{H_b}(\kp)$ }

Let us consider the Hamiltonian  $H_b$ 
\begin{equation}
  H_b = \frac{1}{2}\,\bigl(p_x^2 + p_y^2\bigr)  + \frac{1}{2}\,\al^2\,\bigl(4x^2 + y^2\bigr)  +  \frac{k_2}{y^2}   +  k_3 x
\end{equation}  
and denote by  $\la_b$ the following multiplier
$$
   \la_b  = \frac{1}{\mu_b}\,,{\quad}   \mu_b(x,y) = 1 - \kp\,x  \,. 
$$
Then the new $\kp$-dependent Hamiltonian $\wt{H_b}(\kp)$ is thus given by 
\begin{equation}
 \wt{H_b}(\kp) = \frac{1}{2}\,\Bigl(\frac{p_x^2 + p_y^2}{1 - \kp\,x}\Bigr)  + \frac{1}{2}\,\al^2\,\Bigl(\frac{4x^2 + y^2}{1 - \kp\,x}\Bigr)  +  \frac{k_2}{(1 - \kp\,x)\,y^2} 
  +  \frac{k_3x}{(1 - \kp\,x)} \,. 
\end{equation}

\begin{itemize}

\item[(i)] {Cartesian separability }

The H-J equation takes the form 
$$
\frac{1}{2}\,\Bigl(\frac{1}{1 - \k\,x}\Bigr) \Bigl[\Bigl(\fracpd{W}{x}\Bigr)^2 + \Bigl(\fracpd{W}{y}\Bigr)^2\Bigr]  + \frac{1}{2}\,\al^2\,\Bigl(\frac{4x^2 + y^2}{1 - \kp\,x}\Bigr)  +  \frac{k_2}{(1 - \kp\,x)\,y^2}  +  \frac{k_3x}{(1 - \kp\,x)}  = E
$$
so that if we assume that the function $W$ is of the form $W=W_x(x)+W_y(y)$ we can perform a separation of variables and arrive to 
$$
 \Bigl[\Bigl(\fracpd{W_x}{x}\Bigr)^2 +  4\al^2 x^2 +   2 k_3x  + 2 \kp E x\Bigr]  +  
 \Bigl[\Bigl(\fracpd{W_y}{y}\Bigr)^2 +  \al^2 y^2 +  \frac{2 k_2}{y^2} \Bigr]  = 2  E  \,. 
$$
This means that the following two functions 
$$
 J_{b1} = p_x^2 + 4\al^2 x^2 +   2 k_3x  + 2  \kp  x\wt{H_b} \,,{\quad} 
 J_{b2} = p_y^2 +   \al^2 y^2 +  \frac{2 k_2}{y^2} \,, 
$$
 are constants of motion satisfying the following properties 
$$
 (i)\   dJ_{b1}\,\wedge\, dJ_{b2} \ne 0  \,,{\quad} 
 (ii)\   \{J_{b1}\,,\, J_{b2}\} = 0  \,,{\quad} 
 (iii)\    \wt{H_b}(\kp) = \frac{1}{2}\bigl(J_{b1} + J_{b2} \bigr) \,. 
$$

\item[(ii)] {Parabolic separability  } 

If we introduce the following change 
$$ 
 (x,y)\ \to\ (a,b)\,, {\quad}   x = a^2 - b^2 \,,{\quad} y = 2 a b \,, 
$$
then the $\kp$-dependent Hamiltonian becomes 
$$
  \wt{H_b}(\kp) = \frac{1}{2}\,\frac{1}{1 - \kp\,(a^2-b^2)}\Bigl(\frac{p_a^2 + p_b^2}{a^2+b^2}\Bigr)  + \frac{1}{1 - \kp\,(a^2-b^2)}\Bigl[\frac{1}{2}\,\al^2\,\Bigl(\frac{a^6 + b^6}{a^2+b^2}\Bigr)  +  \frac{k_2}{a^2b^2} 
  + k_3 \frac{a^4-b^4}{a^2b^2}\Bigr]
  $$
and the H-J equation 
$$
\frac{1}{2}\,\frac{1}{(a^2+b^2)} \Bigl[\Bigl(\fracpd{W}{a}\Bigr)^2 + \Bigl(\fracpd{W}{b}\Bigr)^2\Bigr]  +  \Bigl[\frac{1}{2}\,m\al^2\,\Bigl(\frac{a^6 + b^6}{a^2+b^2}\Bigr)  +  \frac{k_2}{a^2b^2} 
  + k_3 \frac{a^4-b^4}{a^2b^2}\Bigr]  = \bigl(1 - \kp\,(a^2-b^2)\bigr)E
$$
also admits separation of variables 
$$
  \Bigl[\Bigl(\fracpd{W_a}{a}\Bigr)^2 + \al^2 a^6 + \frac{2 k_2}{a^2}  +  2 k_3 a^4 - 2  (a^2 - \kp a^4) E\Bigr]   + 
\Bigl[\Bigl(\fracpd{W_b}{b}\Bigr)^2  +  \al^2 b^6 + \frac{2 k_2}{b^2}  - 2 k_3 b^4 - 2  (b^2 + \kp b^4) E\Bigr]  = 0  
$$
The result is that the following function  
\begin{eqnarray*}
 J_{b3}  &=& p_a^2 +   \al^2 a^6 + \frac{2 k_2}{a^2}  +  2 k_3 a^4 - 2 (a^2 - \kp a^4) \wt{H_b}(\kp) \,, \cr
    &=& -\,\Bigl[\,p_b^2 +  \al^2 b^6 + \frac{2 k_2}{b^2}  - 2 k_3 b^4- 2 (b^2 + \kp b^4) \wt{H_b}(\kp) \Bigr]  
\end{eqnarray*}
is also a constant of motion.
\end{itemize}

The following proposition summarizes these results. 

\begin{proposicion} 
The $\kp$-dependent Hamiltonian $\wt{H_b}(\kp)$ is H-J separable both  in Cartesian coordinates  $(x,y)$ and in parabolic coordinates  $(a,b)$, and  it is endowed with the following  three quadratic constants of motion 
\begin{eqnarray*}
 J_{b1}  &=& \frac{1}{(1 - \kp\,x)} \Bigl( p_x^2 + \k x p_y^2 +  \al^2  (4x +\kp y^2)  x+ \frac{2 k_2\kp x}{y^2} + 2 k_3x\Bigr)   \,, \cr
 J_{b2} &=& p_y^2 + \al^2 y^2 +  \frac{2 k_2}{y^2}   \,, \cr
 J_{b3}  &=&  (x p_y  - y p_x ) p_y - \kp\,\frac{y^2(p_x^2+p_y^2)}{4(1 - \kp\,x)}
 + \frac{1}{(1 - \kp\,x)} \biggl(-\, \frac{\al^2}{4} y^2 (4 x + \kp y^2)
 + \frac{k_2}{2 y^2} \bigl(4 x - \kp (4 x^2 + y^2)\bigr) - k_3 \frac{y^2}{2}  \biggr) \,. 
\end{eqnarray*}
\end{proposicion}

\subsection{Hamiltonian  $\wt{H_c}(\kp)$ }

Let us consider the Hamiltonian  $H_c$ 
\begin{equation}
  H_c = \frac{1}{2m}\,\bigl(p_x^2 + p_y^2\bigr)  + \frac{k_1}{\sqrt{x^2 + y^2}}   +  \frac{k_2}{y^2}   + k_3 \frac{x}{y^2 \sqrt{x^2 + y^2}}
\end{equation}  
and denote by  $\la_c$ the following multiplier
$$
 \la_c  = \frac{1}{\mu_c}\,,{\quad}   \mu_c(x,y) =  1 - \frac{\kp}{r} \,. 
$$
Then the new $\kp$-dependent Hamiltonian $\wt{H_c}(\kp)$ is 
\begin{equation}
  \wt{H_c}(\kp) = \frac{1}{2}\, \Bigl(\frac{r}{r - \kp}\Bigr) (p_x^2 + p_y^2)  + 
  \frac{k_1}{r - \kp}   +  \frac{k_2\,r}{(r - \kp)\,y^2}  +  \frac{k_3x}{(1 - \kp\,r)\,y^2} \,. 
\end{equation}

\begin{itemize}

\item[(i)]   {Polar separability }

The H-J equation takes the form 
$$
\frac{1}{2}\, \Bigl(\frac{r}{r - \kp}\Bigr) \Bigl[\Bigl(\fracpd{W}{r}\Bigr)^2 +  \frac{1}{r^2}\Bigl(\fracpd{W}{\phi}\Bigr)^2\Bigr]  +  \frac{k_1}{r - \kp}  +  \frac{k_2\,r}{(r - \kp)\,r\sin^2\phi}  +  \frac{k_3\cos\phi}{(r - \kp)\,r\sin^2\phi}   =  E
$$
so that if we assume that $W$ is of the form $W=W_r(r)+W_\phi(\phi)$ we can perform a separation of variables and rewrite the equation as the sum of two one-variable summands
$$
  \Bigl[r^2\Bigl(\fracpd{W_r}{r}\Bigr)^2 + 2 k_1r  -2 \, r (r - \kp) E\Bigr]   +
\Bigl[\Bigl(\fracpd{W_\phi}{\phi}\Bigr)^2 +  \frac{2 k_2}{\sin^2\phi}  + 2 k_3\frac{\cos\phi}{\sin^2\phi} \Bigr] =  0
$$
so that the following function
$$
 J_{c2} = p_\phi^2 +  \frac{2 k_2}{\sin^2\phi}  + 2 k_3\frac{\cos\phi}{\sin^2\phi}
$$
is a constant of motion. 

\item[(ii)]   {Parabolic separability }

The Hamiltonian $\wt{H_c}$ takes the following form in parabolic coordinates 
$$
  \wt{H_c}(\kp) = \frac{1}{2}\,\frac{1}{(a^2+b^2) - \kp}\Bigl(p_a^2 + p_b^2\Bigr)  + \frac{1}{(a^2+b^2) - \kp}\Bigl[2 k_1  +  k_2 \frac{a^2+b^2}{a^2b^2} 
  + k_3 \frac{a^2-b^2}{a^2b^2} \Bigr]
  $$
so that the H-J equation becomes 
$$
 \Bigl[\Bigl(\fracpd{W}{a}\Bigr)^2 + \Bigl(\fracpd{W}{b}\Bigr)^2\Bigr]    
  +  2 \Bigl[2 k_1  +  k_2 \frac{a^2+b^2}{a^2b^2} 
  + k_3 \frac{a^2-b^2}{a^2b^2} \Bigr] = 2  ((a^2+b^2) - \kp)E
$$
and it leads to 
\begin{eqnarray*}
  \Bigl[\Bigl(\fracpd{W_a}{a}\Bigr)^2 + 2 k_1 + \frac{2 k_2}{a^2}  - \frac{2 k_3}{a^2}  - 2 a^2 E \Bigr]   &=&   -\,\kp E + K    \cr
\Bigl[\Bigl(\fracpd{W_b}{b}\Bigr)^2 + 2 k_1 + \frac{2 k_2}{b^2} + \frac{2 m k_3}{b^2}   - 2 b^2 E \Bigr]  &=& -\,\kp E - K  
\end{eqnarray*}
so that the following two functions 
\begin{eqnarray*}
 J_{c3a}  &=& p_a^2 + 2 k_1 + \frac{2 k_2}{a^2}  - \frac{2 k_3}{a^2}  - 2 a^2  \wt{H_c}(\kp) \,, \cr
 J_{c3b}  &=& p_b^2 + 2 k_1 + \frac{2 k_2}{b^2} + \frac{2 k_3}{b^2}  - 2 b^2  \wt{H_c}(\kp)  \,,
\end{eqnarray*}
are two constants of motion representing two different $\kp$-deformations of $I_{c3}$ 
$$
 (i)\  dJ_{c3a}\,\wedge\, dJ_{c3b}\ne 0 \,,{\quad} (ii)\    \{J_{c3a}\,,\, J_{c3b}\} = 0
 \,,{\quad} (iii)\    \lim_{\kp\to 0}J_{c3a}  = \lim_{\kp\to 0}J_{c3b} =I_{c3}  \,. 
$$
Nevertheless, as the only difference between them is a term proportional to $\wt{H_c}(\kp)$ (with $\kp$ as coefficient), we consider as a more appropiate third constant the following function
\begin{eqnarray*}
  J_{c3}  &=&  p_a^2 + 2 k_1 + \frac{2 k_2}{a^2}  - \frac{2 k_3}{a^2} - 2 a^2  \wt{H_c}(\kp) + \kp \wt{H_c}(\kp)    \cr 
   &=& -\,\Bigl[\,p_b^2 + 2 k_1 + \frac{2 k_2}{b^2} + \frac{2 k_3}{b^2}   - 2 b^2  \wt{H_c}(\kp) + \kp \wt{H_c}(\kp) ) \Bigr]  
\end{eqnarray*}
that in a more detailed way is as follows 
\begin{eqnarray*}
  J_{c3}  &=&   \frac{1}{a^2 + b^2 - \kp} \biggl[(a^2p_b^2 - b^2p_a^2) 
  + \frac{\kp}{2}\, (p_a^2-p_b^2)  +  2 k_1\,(a^2 - b^2)   \cr 
 && +  \frac{k_2}{a^2 b^2} \Bigr(2 (a^4 - b^4) - \kp(a^2-b^2)  \Bigr) 
      +  \frac{k_3}{a^2 b^2} \Bigr(2 (a^4 + b^4) - \kp(a^2+b^2) \Bigr)  \biggr] \,. 
\end{eqnarray*}
\end{itemize}

\begin{proposicion} 
The $\kp$-dependent Hamiltonian  $\wt{H_c}(\kp)$ is H-J separable in polar coordinates  $(r,\phi)$ and parabolic coordinates  $(a,b)$ and  it is endowed with the Hamiltonian as the first constant, that is $J_{c1}=\wt{H_c}(\kp)$, and the following  two additional quadratic constants of motion 
\begin{eqnarray*}
   J_{c2} &=&  (x p_y - y p_x)^2 + \frac{2 k_2 x^2}{y^2} +
  \frac{2 k_3 x \sqrt{x^2 + y^2}}{y^2}   \,, \cr
   J_{c3}  &= &\frac{1}{r - \kp} \biggl[r\,(xp_y - yp_x)  p_y  + \frac{\kp}{2}\, (xp_x^2  - x p_y^2  + 2 y p_x p_y )    +    k_1\,x\cr 
&{}&  {\hskip30pt}+\, \frac{k_2x}{y^2} (2 r - \kp) 
     + \frac{k_3}{y^2} \bigr(2 x^2 + y^2 -\kp\, r\bigr)  \biggr]  \,. 
\end{eqnarray*}
\end{proposicion}

The function $J_{c2}$ it is $\kp$-independent and it coincides with the original constant  $I_{c2}$ (the same situation we found in the case (a)  with $J_{a3}$). 
Concernig $J_{c3}$, it is clear that it satisfies the limit $J_{c3} \to I_{c3}$ when 
${\kp\to 0}$.

\subsection{Hamiltonian  $\wt{H_d}(\kp)$ }

Let us consider the Hamiltonian  $H_d$ 
\begin{equation}
  H_d = \frac{1}{2}\,\bigl(p_x^2 + p_y^2\bigr)  + \frac{k_1}{r}   
  +  k_2 \frac{\,\sqrt{r + x}\,}{r}  +  k_3 \frac{\,\sqrt{r - x}\,}{r}   
  \,,{\quad}  r^2 = x^2+y^2\,,
\end{equation}
and denote by  $\la_d$ the following multiplier
$$
 \la_d  = \frac{1}{\mu_d}\,,{\quad}   \mu_d(x,y) = 1 - \frac{\kp}{r} \,. 
$$
Thus, the $\kp$-dependent Hamiltonian we will study is 
\begin{equation}
  \wt{H_d}(\kp) = \frac{1}{2}\, \Bigl(\frac{r}{r - \kp}\Bigr) (p_x^2 + p_y^2)  + 
  \frac{k_1}{r - \kp}   +  \frac{k_2\,\sqrt{r+x}}{r - \kp}  
  +  \frac{k_3\sqrt{r-x}}{r - \kp} \,. 
\end{equation}

\begin{itemize}

\item[(i)]   {Parabolic separability I}

The Hamiltoniano $\wt{H_d}(\kp)$ takes the following form when written in parabolic coordinates
$$
  \wt{H_d}(\kp) = \frac{1}{2}\,\frac{1}{(a^2+b^2) - \kp}\Bigl(p_a^2 + p_b^2\Bigr)  + \frac{1}{(a^2+b^2) - \kp}\Bigl( k_1  +  k_2 a  + k_3 b \Bigr)
  $$
so that the corresponding  H-J Equation 
$$
\frac{1}{2}\,\frac{1}{(a^2+b^2) - \kp} \Bigl[\Bigl(\fracpd{W}{a}\Bigr)^2 + \Bigl(\fracpd{W}{b}\Bigr)^2\Bigr]  + \frac{1}{(a^2+b^2) - \kp}\Bigl( k_1  +  k_2 a  + k_3 b \Bigr) = E
$$
admits separation of variables and it reduces to 
$$
 \Bigl[\Bigl(\fracpd{W_a}{a}\Bigr)^2 +  k_1 + 2 k_2 a - 2 a^2 E \Bigr]   + \Bigl[\Bigl(\fracpd{W_b}{b}\Bigr)^2  +  k_1 + 2 k_3 b - 2 b^2 E \Bigr]    = - 2 \kp E \,. 
$$
Thus,  the following two functions
\begin{eqnarray*}
 J_{d2a}  &=&   p_a^2  +  k_1 + 2 k_2 a - 2 a^2  \wt{H_d}(\kp) \,, \cr
 J_{d2b}  &=&   p_b^2  +  k_1 + 2 k_3 b - 2 b^2  \wt{H_d}(\kp)  \,, 
\end{eqnarray*}
that written with more detail are as follows 
\begin{eqnarray*}
  J_{d2a}  &=&  \frac{1}{(a^2+b^2) - \kp}\Bigl[a^2p_b^2 - b^2 p_a^2 + \kp p_a^2   +  k_1(a^2-b^2+\kp) -  2 k_2 a (b^2 - \kp)  +  2 k_3 a^2 b   \Bigr] \cr 
  J_{d2b} &=&  \frac{1}{(a^2+b^2) - \kp}\Bigl[a^2p_b^2 - b^2 p_a^2 - \kp p_b^2   +  k_1(a^2-b^2-\kp) -  2 k_2  a b^2  +  2 k_3 b (a^2 - \kp)   \Bigr]
\end{eqnarray*}
are two independent constants of motion.  
  Nevertheless as the only difference between them is just a term proportional to the Hamiltonian, that is $J_{d2b} - J_{d2a} =  2\,  \kp \,\wt{H_d}(\kp)$, we consider more convenient to choose the following function
$$
  J_{d2}  = J_{d2b}-\kp \,\wt{H_d}(\kp) = J_{d2a}+\kp \,\wt{H_d}(\kp) 
 $$
that takes the form 
$$
 J_{d2} =  \frac{1}{a^2 + b^2 - \kp}\Bigl((a^2p_b^2-b^2p_a^2) 
  + \frac{\kp}{2}\, (p_a^2-p_b^2) \Bigr) + \frac{1}{a^2 + b^2 - \kp}
 \Bigl( k_1\,(a^2 - b^2) - k_2\,a  (2 b^2 - \kp) + k_3\,b (2 a^2 - \kp) \Bigr) 
$$
as the integral representing the $\kp$-dependent version of $I_{d2}$. 

\item[(ii)]   {Parabolic separability II}

We can introduce a new system of parabolic coordinates by rotating the original system 
$$
 (a,b)\to(\al,\be) \  ;\  a = \frac{1}{\sqrt{2}}(\al+\be), \  b = \frac{1}{\sqrt{2}}(\al-\be) \,,
$$
in such a way that the Hamiltonian $\wt{H_d}(\kp)$, when written in this new system, it takes the following form  
$$
  \wt{H_d}(\kp) = \frac{1}{2}\,\frac{1}{(\al^2+\be^2) - \kp}\Bigl(p_\al^2 + p_\be^2\Bigr)  + \frac{1}{(\al^2+\be^2) - \kp}\Bigl[ k_1  +  k_2 \frac{1}{\sqrt{2}}(\al+\be)  + k_3 \frac{1}{\sqrt{2}}(\al-\be)\Bigr]   \,. 
$$
The H-J equation  
$\wt{H_d}(\partial W/\partial\alpha,\partial W/\partial\alpha,\alpha,\beta)=E$  
is separable as well, and it leads to the following two functions 
\begin{eqnarray*}
 J_{d3a}  &=& p_\al^2  +  k_1 + \sqrt{2} k_2 \al  + \sqrt{2} k_3 \al - 2 \al^2  \wt{H_d}(\kp) \,,   \cr
 J_{d3b}  &=& p_\be^2  +  k_1 + \sqrt{2} k_2 \be  - \sqrt{2} k_3 \be - 2 \be^2  \wt{H_d}(\kp) \,,  
\end{eqnarray*}
that recovering the coordinates $(a,b)$ appear as follows 
$$
  J_{d3a}  =  \frac{1}{(a^2+b^2) - \kp}\Bigl[(ap_b - b p_a)(a p_a-b p_b) - \frac{\kp}{2} (p_a + p_b)^2  - k_1(2a b +\kp) +  k_2 (a+b)  (b^2- a b - \kp)  +  k_3  (a+b) (a^2- a b - \kp) \Bigr]
$$
$$
  J_{d3b} =  \frac{1}{(a^2+b^2) - \kp}\Bigl[(ap_b - b p_a)(a p_a-b p_b) + \frac{\kp}{2}(p_a - p_b)^2  - k_1(2a b -\kp) -  k_2 (a-b)  (b^2+a b - \kp)  +  k_3  (a-b) (a^2+ a b - \kp) \Bigr]
$$

This situation is similar to the previous one, that is, they are independent but the only difference between them is $\kp$ times de Hamiltonian, that is $J_{d3b} - J_{d3a} =  2\,  \kp \,\wt{H_d}(\kp)$;  therefore we choose the following function $J_{d3}$ as the integral of motion associated to the system $(\alpha,\beta)$
$$   
 J_{d3}  =  \frac{1}{a^2 + b^2 - \kp}\Bigl((ap_b-bp_a)(ap_a-bp_b) - \kp\, p_a p_b  \Bigr) + \frac{1}{a^2 + b^2 - \kp}\Bigl( - 2 k_1\,a b  - k_2\,b  (a^2 - b^2 +\kp) + k_3\,a  (a^2 - b^2 -\kp) \Bigr) \,. 
$$
\end{itemize}

\begin{proposicion} 
The $\kp$-dependent Hamiltonian  $\wt{H_d}(\kp)$ is H-J separable in two different systems of parabolic coordinates, the origial system $(a,b)$ and the rotated system $(\al,\be)$, and it is endowed with three quadratic constants of motion; the Hamiltonian as the first constant, that is $J_{d1}=\wt{H_d}(\kp)$, and the following  two additional integrals
\begin{eqnarray*}
 J_{d2} &=&   \frac{1}{r - \kp} \Bigl[r\,(xp_y - yp_x)  p_y  + \frac{\kp}{2}\, \Bigl(xp_x^2  - x p_y^2  + 2 y p_x p_y \Bigr)   +    k_1\, x \cr 
&{}&{\hskip30pt}  -\,  k_2 \Bigl( y \sqrt{r-x} - \kp\sqrt{r+x} \Bigr)  
+ k_3 \Bigl( y \sqrt{r+x} - \kp\sqrt{r-x} \Bigr)  \,\Bigr]\,, \cr
 J_{d3}  &= & \frac{1}{r - \kp} \Bigl[r\,(xp_y - yp_x)  p_x  + \frac{\kp}{2}\, \Bigl(yp_x^2  - y p_y^2  - 2 x p_x p_y \Bigr)    -    k_1\,y\cr 
&{}&{\hskip30pt}  -\,  k_2 (x + \kp)\sqrt{r-x}     +  k_3 (x-\kp)\sqrt{r+x} \, \Bigr]  \,. 
\end{eqnarray*}
\end{proposicion}

\subsection{Comments }

The previous paragraphs can be considered as rather  technical;  so now we comment some of the characteristics of these new families of Hamiltonians and we analyze to what physics systems they seem to correspond.

\begin{itemize}

\item[(i)] All these ``new Hamiltonians" are of the form of the original Hamiltonian multiplied by a factor. How was this factor found ? The answer is that every $\la_r$ is obtained  by imposing separability in two  different systems of coordinates and this property determines  $\la_r$ with the only ambiguity of the appropriate position and sign of the parameter $\kp$. 

It is clear that the factors $\la_r$ introduce a change in the geometry of the configuration space.  The new deformed Hamiltonians $\wt{H_r}(\kp)$ describe dynamics in Riemannian manifolds (a particle moving in a curved space); the negative sign in front of $\kp$ has been chosen in order that the sign of $\kp$ coincides with the sign of the curvature.

\item[(ii)] The expressions of the four Hamiltonians  $\wt{H_r}(\kp)$ can be considered as rather involved mainly because the nonlinearity affects to both the kinetic term and the potential.  
Nevertheless the fundamental point is that, in spite of their complex aspect, they are directly related with the two fundamental superintegrable systems, that is,  harmonic oscillators and Kepler problems.

The two first Hamiltonians, $\wt{H_a}(\kp)$ and $\wt{H_b}(\kp)$, are related with the harmonic oscillator. More specifically, they are deformations of the nonlinear oscillators $H_a$ and $H_b$.  Therefore $\wt{H_a}(\kp)$ and $\wt{H_b}(\kp)$ describe nonlinear oscillators in non-Euclidean spaces. 

The other two Hamiltonians, $\wt{H_c}(\kp)$ and $\wt{H_d}(\kp)$,  are deformations of the nonlinear Hamiltonians $H_c$ and $H_d$ and therefore they are related with the Kepler problem. For that reason $\wt{H_c}(\kp)$ and $\wt{H_d}(\kp)$ describe two different versions of the Kepler problem  in two different non-Euclidean spaces. 

\end{itemize}

The following section studies these questions with more detail. 

\section{The harmonic oscillator and the Kepler problem on curved spaces }

In differential geometric terms, the four Hamiltonians $\wt{H_r}(\kp)$, $r=a,b,c,d$, describe dynamics on non-Euclidean spaces. 
The two first Hamiltonians, $\wt{H_a}(\kp)$ or $\wt{H_b}(\kp)$, are related with the harmonic oscillator and the other two, $\wt{H_c}(\kp)$ or $\wt{H_d}(\kp)$, with the Kepler problem. Now in this section we analyze these two important $\kp$-dependent systems from a geometric perspective.

\subsection{The harmonic oscillator on curved spaces }

The following two equations 
\begin{equation}  
\frac{d^2x}{dt^2}+\frac{\kp\, x}{1-\kp x^2}\left(\frac{dx}{dt}\right)^2 + \frac{\alpha^2x}{1-\kp x^2} = 0  \,, 
\end{equation}
\begin{equation} 
 \frac{d^2x}{dt^2} - \frac{\kp\,x}{1-\kp x^2}\left(\frac{dx}{dt}\right)^2 
+ \frac{\alpha^2x}{(1- \kp x^2)^3} = 0  \,, 
\end{equation}
represent two one-dimensional nonlinear oscillators that can be considered as arising from the Lagrangians 
$$
  L_1= \frac{1}{2} \frac{v_x^2}{1-\kp x^2}   - \frac{1}{2} \frac{\alpha^2\,x^2}{1-\kp x^2}  
{\quad} {\rm and}{\quad}
  L_2 = \frac{1}{2}(1- \kp  x^2)\,v_x^2   - \frac{1}{2}\frac{\alpha^2\,x^2}{1-\kp x^2} \,, 
$$
with associated Hamiltonians 
\begin{equation}  
  H_1 = \frac{1}{2} (1-\kp x^2)p_x^2  +   \frac{1}2\frac{\alpha^2\,x^2}{1-\kp x^2}  \,,  
\end{equation}    
and 
\begin{equation}    
  H_2 = \frac{1}{2}\frac{p_x^2}{1- \kp x^2}  +   \frac{1}{2}\frac{\alpha^2\,x^2}{1- \kp x^2}  \,. 
\end{equation}  
That is, the two potentials are the same (see Figure 1) but the kinetic terms are different  (these two oscillators are studied in \cite{MatLak74}-- \cite{CaRaSaAnPh07a}).

The two-dimensional versions of these Lagrangians are 
\begin{eqnarray*}  
 L_1(\kp)  &=&  \frac{1}{2}\,\Biggl(\frac{v_r^2}{1 - \kp\,r^2} + r^2v_\phi^2 \,\Biggr) - \frac{\al^2}{2}\Bigl(\frac{r^2}{1 - \kp\,r^2} \Bigr)    \cr 
   &=&  \frac{1}{2}\,\Bigl(\frac{1}{1 - \kp\,r^2} \Bigr)
 \Bigl[\,v_x^2 + v_y^2 - \kp\,(x v_y - y v_x)^2 \,\Bigr]
 - \frac{\al^2}{2}\Bigl(\frac{x^2 + y^2}{1 - \kp\,r^2} \Bigr)\,,
\end{eqnarray*}  
and
$$
 L_2(\kp)  =  \frac{1}{2}(1 - \kp\,r^2)\,(v_x^2 + v_y^2) - \frac{\al^2}{2}\Bigl(\frac{x^2 + y^2}{1 - \kp\,r^2}  \Bigr)\,.  
$$
 It is known that a symmetric bilinear form in the velocities $(v_x,v_y)$ can be considered as associated to a two-dimensional metric $ds^2$ in $\IR^2$. 
 In this particular case, the kinetic term in the Lagrangian $L_1(\kp)$ considered as a bilinear form determines the following $\kp$-dependent metric
$$
 ds_\kp^2 = \Bigl(\frac{1}{1 - \kp\,r^2}\Bigr)\,  \Bigl[\,(1 - \kp\,y^2)\,dx^2 
 + (1 - \kp\,x^2)\,dy^2 + 2 \kp\,x y \,dx \,dy\,\Bigr]\,.
$$
The second Lagrangian $L_2(\kp)$ represents the harmonic oscillator in a two-dimensional space with a metric $ds_\kp^2$ conformally flat 
$$
  ds_\kp^2 = (1 - \kp\,r^2)(dx^2 + dy^2)  \,. 
$$

The two-dimensional versions of these Hamiltonian are 
\begin{equation}
 H_1(\kp) = \frac{1}{2}\,\Bigl((1 - \kp\,r^2)(p_x^2 + p_y^2) + \kp J^2\Bigr)  + \frac{1}{2}\,\al^2\,\Bigl(\frac{x^2 + y^2}{1 - \kp\,r^2}\Bigr)  \,,  
\end{equation}  
and
\begin{equation}
 H_2(\kp) = \frac{1}{2}\,\Bigl(\frac{p_x^2 + p_y^2}{1 - \kp\,r^2}\Bigr)  + \frac{1}{2}\,\al^2\,\Bigl(\frac{x^2 + y^2}{1 - \kp\,r^2}\Bigr) \,, 
\end{equation}
where $J$ denotes the angular momentum. They satisfy the correct Euclidean limit 
$$
 \lim_{\kp\to 0} H_1(\kp) = \lim_{\kp\to 0} H_2(\kp) =
 \frac{1}{2}\,\bigl(p_x^2 + p_y^2\bigr)  + \frac{1}{2}\,\al^2\bigl(x^2 + y^2\bigr)   \,, 
$$
and  represent two different harmonic oscillators in two-dimensional curved spaces.

\begin{itemize}

\item[(i)]
The Hamiltonian $H_1(\kp)$, that has been studied in Refs. \cite{CaRaSS04, CaRaSaAnPh07b, CaRaSaIntJ11, CaRaSaJPh12} (although in some cases with a trigonometric-hyperbolic  notation), represents the harmonic oscillator in a space of constant curvature $\kp$ (sphere $S_\kp^2$ with $\kp>0$, and Hiperbolic plane $H_\kp^2$ with $\kp<0$).  
We note that the kinetic term includes not only the factor $(1 - \kp\,r^2)$ but also a contribution of the angular momentum $J$ with the curvature $\kp$ as coefficient. 

\item[(ii)] 
The second Hamiltonian $H_2(\kp)$, that is just $\wt{H_a}(\kp)$ with $k_2=k_3=0$, represents an harmonic oscillator in a curved space (two-dimensional space of nonconstant curvature)  \cite{BurgosPhysd08}--\cite{BurgosSigma11}. 
In two dimensions the tensor $R_{abcd}$ only has one independent component which can be taken $R_{1212}$ 
$$
 R_{1212} =  \frac{1}{2}\,\bigl(\partial_2\partial_1g_{21} - \partial_2^2g_{11} + \partial_1\partial_2g_{12} - \partial_1^2g_{22} \bigr) 
 - g_{ef}\bigl( \Ga_{11}^e\Ga_{22}^f   -   \Ga_{12}^e\Ga_{21}^f  \bigr) \,. 
$$
The result is 
$$
 R_{1212} =  \frac{2\,\kp}{1 - \kp\,r^2}
$$
and the scalar Gaussian curvature  is given by
$$
  K = \frac{R_{1212}}{\det[g]} =  \frac{2\,\kp}{(1 - \kp\,r^2)^3}  \,. 
$$
\end{itemize}

\subsection{The Kepler problem on curved spaces }
 
The following two $\kp$-dependent Hamiltonians 
\begin{equation}  
 H_{K1}(\kp) = \frac{1}{2}\,\Bigl((1 - \kp\,r^2)(p_x^2 + p_y^2) + \kp J^2\Bigr)  
 - g\Bigl(\frac{\sqrt{1 - \kp\,r^2}}{r}\,\Bigr)  
\end{equation}
and
\begin{equation} 
 H_{K2}(\kp) =  \frac{1}{2}\, \Bigl(\frac{r}{r - \kp}\Bigr) (p_x^2 + p_y^2)  - 
  \frac{g}{r - \kp}  
\end{equation}  
represent two different versions of the Kepler problem on curved spaces. 
They satisfy the correct Euclidean limit 
$$
 \lim_{\kp\to 0} H_{K1}(\kp) = \lim_{\kp\to 0} H_{K2}(\kp) =
 \frac{1}{2}\,\bigl(p_x^2 + p_y^2\bigr) -  \frac{g}{r}  \,. 
$$
 
\begin{itemize}

\item[(i)]
The Hamiltonian $H_{K1}(\kp)$, that has been studied in Refs. \cite{CaRaSaJmp05, CaRaSaJpa07}, represents the Kepler problem in the three spaces of constant curvature $\kp$; that is,  sphere $S_\kp^2$ ($\kp>0$), Euclidean plane $E^2$ ($\kp=0$), and Hiperbolic plane $H_\kp^2$ ($\kp<0$). Also in this case, when $\kp>0$,  the dynamics is restricted to the interior of the bounded region $x^2+y^2<1/\kp$. 

\item[(ii)]
The Hamiltonian $H_{K2}(\kp)$ is just the Hamiltonian of the Kepler problem in $\wt{H_c}(\kp)$ or $\wt{H_d}(\kp)$. In this case if $\kp$ is negative then the Hamilltonian is well defined  for all the values $\kp<0$,  but when $\kp>0$ then the dynamics  is only well defined in the region $r>\kp$ in such a way that when $\kp\to 0$ then the dynamics is defined in the whole space. 
This system is endowed with the following two integrals of motion 
\begin{eqnarray*}
 J_{2} &=&    \frac{1}{r - \kp} \Bigl[r\,(xp_y - yp_x)  p_y   -   g\, x \,\Bigr] 
 + \frac{\kp}{2(r - \kp) }\, \Bigl(xp_x^2  - x p_y^2  + 2 y p_x p_y \Bigr)   \, \cr 
 J_{3}  &= & \frac{1}{r - \kp} \Bigl[r\,(xp_y - yp_x)  p_x   +  g\, y \,\Bigr] 
 + \frac{\kp}{2(r - \kp) }\, \Bigl(yp_x^2  - y p_y^2  - 2 x p_x p_y  \Bigr) 
\end{eqnarray*}
that represent the $\kp$-dependent version of the (two-dimensional) Runge-Lenz vector.  In fact, it can be verified that their Poisson bracket is given by 
$$
\bigl\{J_{2}\,, J_{3}\bigr\} = 2 (x p_y - y p_x)\,H_{K2}(\kp)
$$
that is the same relation characterizing the Runge-Lenz vector in the Euclidean case. 
\end{itemize}

The metric $ds_\kp^2$, that is also conformal,  is given by 
$$
  ds_\kp^2 = \Bigl(1-\frac{\kp}{r}\Bigr)(dx^2 + dy^2) \,. 
$$
We note the factor $(1-\kp/r)$ shows a certain similarity with the coefficient (related with the singularity) in the Schwarzschild metric. 
The curvature tensor $R_{1212}$ takes the value 
$$
 R_{1212} =  \frac{\kp}{2 \,r^2 \,(r - \kp)}
$$
and the Gaussian curvature (scalar) is given by
$$
  K = \frac{R_{1212}}{\det[g]} =  \frac{\kp}{2\,(r - \kp)^3} \,. 
$$

\section{Superintegrabilty with higher orden constants of motion } 

We have proved that the two Hamiltonians $\wt{H_a}(\kp)$ (related to the harmonic oscillator) and  $\wt{H_c}(\kp)$ (related to the Kepler problem) are superintegrable with quadratic constants of motion. 
Now, in this section, we will prove that they can be considered as particular cases of a more general situation; that is, they admit superintegrable generalizations that are separable but in only one system of coordinates (two quadratic constants)  and are endowed with an additional constant of higher order. 

At this point (as a previous comment to the next two subsections) we notice that a Hamiltonian of the form 
$$
   H(\k)  = \frac{1}{2}\,\Bigl(\frac{1}{1 - \k\,r^2}\Bigr) \,\Bigl(\, p_r^2 +  \frac{p_{\phi}^2}{r^2}\,\Bigr)   +  \frac{1}{2}\,\al^2\,\Bigl(\frac{A(r)}{1 - \k\,r^2}\Bigr)  
   + \frac{1}{2}\,\frac{1}{1 - \k\,r^2}\,\frac{B(\phi)}{r^2}   \,, 
$$
where $A(r)$ and $B(\phi)$ are arbitrary functions, is H-J separable in polar coordinates $(r,\phi)$ and  it is endowed with the following constant of motion 
$$
 J = p_\phi^2 +  B(\phi)  
    = -\,\bigl[\, r^2p_r^2  + \al^2 r^2 A(r) - 2  r^2  (1 - \k\,r^2) H(\kp) \,\bigr]   \,. 
$$
This property is true for all the values of the parameter $\k$.

\subsection{A $\kp$-dependent Hamiltonian related with the harmonic oscillator and the TTW system} 

The following potential 
\begin{equation}
  V_{ttw}(r,\phi)  =  {\smallonehalf}\,{\om_0}^2 r^2 +  \frac{1}{2\,r^2}\,\Bigl(   \frac{\alpha}{\cos^2(m\phi)}  +  \frac{\beta}{\sin^2(m\phi)} \Bigr) \,,   
\label{VttwE2}
\end{equation}
was  studied by Tremblay, Turbiner, and Winternitz \cite{TTW09}--\cite{TTW10} and also by other authors \cite{Qu10a}--\cite{Ra14JPaTTWk}. 
In the general $m\ne 1$ case  it  is only separable in polar coordinates $(r,\phi)$ ($m$ must be an integer or rational number) and therefore the third integral is not quadratic but a polynomial  of higher order in the momenta (the degree of the polynomial depends of the value of $m$).

Let us  consider the $\kp$-dependent  Hamiltonian 
\begin{equation}
 \wt{H_a}_m(\kp)  = \frac{1}{2}\, \Bigl(\frac{1}{1 - \kp\,r^2}\Bigr) \,\Bigl(\, p_r^2 +  \frac{p_{\phi}^2}{r^2}\,\Bigr) + U_{am}(r,\phi) \,,{\quad}  
  U_{am}(r,\phi)  =  \frac{1}{2}\al^2\Bigl(\frac{r^2}{1 - \kp\,r^2}\Bigr)  
  + \frac{1}{2}\frac{F_m(\phi)}{(1 - \kp\,r^2)\,r^2}   \,, 
\label{HPWk}
\end{equation}
where $F_m(\phi)$ denotes the following angular function 
$$
  F_m(\phi) =  \frac{k_a}{\cos^2(m\phi)}  +  \frac{k_b}{\sin^2(m\phi)} \label{Fm}
$$
and $k_a$ and $k_b$ are arbitrary constants.
It represents a generalization of the Hamiltonian $\wt{H_a}(\kp)$ in the sense that if $m=1$ then we have $\wt{H_a}_1(\kp)=\wt{H_a}(\kp)$. It is clear that this more general Hamiltonian is separable but only in polar $(r,\phi)$ coordinates. 
Therefore it is integrable with the Hamiltonian itself as the first integral and a second quadratic constant of motion $J_2$ associated to the Liouville integrability 
\begin{eqnarray*}  
 J_1(\k) &=& \Bigl(\frac{1}{1 - \kp\,r^2}\Bigr) \,\Bigl(\, p_r^2 +  \frac{p_{\phi}^2}{r^2}\,\Bigr) + \al^2\Bigl(\frac{r^2}{1 - \kp\,r^2}\Bigr)  + \frac{F_m(\phi)}{(1 - \kp\,r^2)r^2} \,,   \cr 
 J_2 &=& p_\phi^2 +  F_m(\phi)  \,. 
\end{eqnarray*}

 Let us denote by $A_r$ and $B_\phi$ be the  complex functions   $A_r = A_{r1} + i\, A_{r2}$ and  $B_{\phi} = B_{\phi 1} + i\, B_{\phi 2}$ with real and imaginary parts, $A_{r a}$ and $B_{\phi a}$, $a=1,2$, be defined as 
$$
 A_{r1} =  \frac{2}{r}\,p_r\,\sqrt{J_2} \,,{\qquad}
 A_{r2} = \frac{1}{(1 - \kp\,r^2)\,r^2} \Bigl(r^2\, p_r^2  + \al^2\, r^4 - (1 - 2\kp\, r^2) J_2 \Bigr) = J_1 - \frac{2}{r^2} \,J_2\,,  
$$
and 
$$
 B_{\phi 1} =   J_2\cos(2m\phi) + (k_b - k_a)  \,,{\qquad}
 B_{\phi 2} =   p_\phi\,\sqrt{J_2}\, \sin(2m\phi) \,.
$$ 

First, let us comment  that the moduli of these two complex functions, that are constant of  motion of fourth order in the momenta,  are given by
\begin{eqnarray*}
  \mid A_r \mid^2  &=& J_1^2 - 4 \kp J_1 J_2  - 4\al^2 J_2 \cr
  \mid B_\phi \mid^2 &=&  J_2^2 - 2(k_a+k_b) J_2 + (k_b-k_a)^2
\end{eqnarray*}

The  Poisson bracket of the function $A_{r1}$ with $\wt{H_a}_m(\kp)$  [time-derivative]  is proportional to $A_{r2}$ and the time-derivative of the $A_{r2}$ is proportional to $A_{r1}$ but with the opposite sign 
$$
 \bigl\{A_{r1}\,, \wt{H_a}_m(\kp)\bigr\} =  -\, 2\,\lambda_{\kp}\,A_{r2} \,,{\quad}
 \bigl\{A_{r2}\,, \wt{H_a}_m(\kp)\bigr\} =   2\,\lambda_{\kp}\,A_{r1}  \,,
$$
and this property is also true for the  functions $B_{\phi 1}$ and $B_{\phi 2}$ 
$$
 \bigl\{B_{\phi 1}\,, \wt{H_a}_m(\kp)\bigr\} =  -2\,m\,\lambda_{\kp}\,B_{\phi 2} \,,{\quad}
 \bigl\{B_{\phi 2}\,, \wt{H_a}_m(\kp)\bigr\} =   2\,m\,\lambda_{\kp}\,B_{\phi 1}  \,,{\quad} 
$$
where the common factor ${\lambda_\kp}$ takes the value 
$$
 {\lambda_\k} = \frac{1}{(1 - \kp\,r^2)\,r^2}\,\sqrt{J_2}  \,,{\quad}
  {\lambda_0} = \frac{1}{r^2}\,\sqrt{J_2}   \,. 
$$
Therefore, the time-evolution of the complex functions $A_r$ and $B_\phi$ is given by
$$
 \frac{d}{d t}\,A_r  =  i\, 2\,  {\lambda_\kp}\,A_r  \,,{\quad}
 \frac{d}{d t}\,B_\phi  =   i\, 2\,   m\,{\lambda_\kp}\,B_\phi   \,.
$$
Thus if we denote by $K_m$ the complex function $A_r^{m}  \, B_\phi^{*}$ then we have
\begin{eqnarray*}
  \frac{d}{dt}\,K_m &=&  \frac{d}{dt}\,\Bigl( A_r^{m}  \, B_\phi^{*} \Bigr) 
  = m\,A_r^{(m-1)} \, \dot{A_r}\,B_\phi^{*} +A_r^{m}\,\dot{B_\phi}^{*}   \cr
  &=& A_r^{(m-1)} \,\biggl( \, m\, (i\, 2\,  {\lambda_\kp}\,A_r )\,B_\phi^{*}  
   +  A_r\,(-\,  i\, 2\,   m\,{\lambda_\kp}\,B_\phi^{*}) \,\biggr) =  0  \,.
\end{eqnarray*}

We summarize this result in the following proposition. 

\begin{proposicion}    \label{prop5} 
The $\kp$-dependent Hamiltonian $ \wt{H_a}_m(\kp)$
\begin{equation}
 \wt{H_a}_m(\kp)  = \frac{1}{2}\, \Bigl(\frac{1}{1 - \kp\,r^2}\Bigr) \,\Bigl(\, p_r^2 +  \frac{p_{\phi}^2}{r^2}\,\Bigr)   +  \frac{1}{2}\al^2\Bigl(\frac{r^2}{1 - \kp\,r^2}\Bigr)    +  \frac{1}{2}\frac{F_m(\phi)}{(1 - \kp\,r^2)\,r^2}   
\label{HamF}
\end{equation}
representing a generalization of the Hamiltonian  $ \wt{H_a}(\kp)$, and also a $\kp$-deformation of the Euclidean TTW Hamiltonian $H_{ttw}$,  is superintegrable with two quadratic constants of motion and a third constant of motion of higher order. 

\begin{itemize}
\item[(i)]  $\wt{H_a}_m(\kp)$  is separable in polar coordinates  $(r,\phi)$ and it possesses therefore two quadratic constants of motion associated to the Liouville integrability 
\begin{eqnarray*}  
 J_1(\k) &=& \Bigl(\frac{1}{1 - \kp\,r^2}\Bigr) \,\Bigl(\, p_r^2 +  \frac{p_{\phi}^2}{r^2}\,\Bigr) + \al^2\Bigl(\frac{r^2}{1 - \kp\,r^2}\Bigr)  + \frac{F_m(\phi)}{(1 - \kp\,r^2)\,r^2}  \,,    \cr 
 J_2 &=& p_\phi^2 +  F_m(\phi)  \,. 
\end{eqnarray*}

\item[(ii)] $\wt{H_a}_m(\kp)$ admits a complex constant of motion $K_m$ defined as
$$
  K_m = A_r^{m} \,B_\phi^{*} \,. 
$$
\end{itemize}
The function $K_m$ can be written as $K_m = J_3 + i J_4$ with $J_3$ and $J_4$ real constants of motion. 
One of them can be chosen as the third fundamental integral of motion. 

\end{proposicion}

\subsection{A $\kp$-dependent Hamiltonian related with the Kepler problen and the PW system }

Let us first  note that the potential $V_c$, that  in polar coordinates becomes 
$$
   V_{c}  =  -\, \frac{g}{r}   +   \frac{1}{r^2} \Bigr(\frac{k_2 }{\sin^2{\phi}}  + \frac{k_3 \cos\phi }{\sin^2{\phi}} \Bigl)  \,,
$$
can also be written as follows 
$$
   V_{c}  =  -\, \frac{g}{r}   +   \frac{1}{r^2}\Bigl(\frac{\alpha} {\cos^2(\phi/2)} +  \frac {\beta} {\sin^2(\phi/2)}\Bigr)   \,,  \ 
   k_2 = 2(\al+\beta)\,, \ k_3=2(\beta-\al) \,.  
$$
Therefore, the angular-dependent functions in the potentials $V_a$ and $V_c$ appear as two particular cases, $m=1$ and $m=1/2$, of the general function $F_m(\phi)$. It seems therefore natural to conjecture that any integrable generalization (or deformation) of the potential $V_a$ must determine a similar generalization (or deformation) of the potential $V_c$.

  In fact there is another interesting system rather similar to  the TTW system but that is related, not with the harmonic oscillator, but with the Kepler problem (hydrogen atom in the quantum case)
\begin{equation}
  V_{pw}(r,\phi)  =  -\, \frac{g}{r}  +  \frac{1}{r^2}\,\Bigl(   \frac{\alpha}{\cos^2(m\phi)} + \frac{\beta}{\sin^2(m\phi)} \Bigr) \,.   \label{VpwE2}
\end{equation}
The first study of the superintegrability of this new potential was presented  by Post and Winternitz \cite{PostWint10} by relating $V_{pw}$ with $V_{ttw}$ making use of  the so-called  coupling constant  metamorphosis transformation (St\"ackel transform) \cite{KaMiPost10}.

Let us now consider the $\kp$-dependent Kepler-related potential 
\begin{equation}
  \wt{H_c}_m(\kp)  =    \frac{1}{2}\, \Bigl(\frac{r}{r - \kp}\Bigr)   \,\Bigl(\, p_r^2 +  \frac{p_{\phi}^2}{r^2}\,\Bigr) + U_{cm}(r,\phi) \,,{\quad}  
U_{cm}(r,\phi)  =  -\,\frac{g}{ r - \kp} + \frac{F_m(\phi)}{( r - \kp)\,r}   \,,
\label{HPWk}
\end{equation}
where $F_m(\phi)$ is the same angular function as in the oscillator $\wt{H_a}_m(\kp)$ (and also in this case $k_a$ and $k_b$ are arbitrary constants).
It represents a generalization of the Hamiltonian $\wt{H_c}(\kp)$ in the sense that if $m=1/2$ then we have $  \wt{H_c}_m(\kp)|_{m=1/2} = \wt{H_c}(\kp)$. 
It is clear that this more general Hamiltonian is separable but only in polar $(r,\phi)$ coordinates. 
Therefore it is integrable with the Hamiltonian itself as the first integral and a second quadratic constant of motion $J_2$ associated to the Liouville integrability 
\begin{eqnarray*}  
 J_1(\k) &=& \Bigl(\frac{r}{r - \kp}\Bigr)   \,\Bigl(\, p_r^2 +  \frac{p_{\phi}^2}{r^2}\,\Bigr)  -\,\frac{2g}{ r - \kp}+ \frac{2 F_m(\phi)}{( r - \kp)\,r}  \,,\cr 
 J_2 &=& p_\phi^2 + 2 F_m(\phi)   \,.
\end{eqnarray*}

We will prove the superintegrability of $\wt{H_c}_m(\kp)$ using as an approach the existence of a complex factorization for the additional constant of motion.  
The method, that is in fact a deformation of the formalism introduced in \cite{Ra13JPaPW} for the superintegrability of the PW system,  is similar to the one presented in the previous section for the Hamiltonian $\wt{H_a}_m(\kp)$ (introducing the appropriate changes).

 Now we denote by $A_r$ and $B_\phi$ be the  complex functions   $A_r = A_{r1} + i\, A_{r2}$ and  $B_{\phi} = B_{\phi 1} + i\, B_{\phi 2}$ with real and imaginary parts, $A_{r a}$ and $B_{\phi a}$, $a=1,2$, be defined as 
$$
 A_{r1} =  p_r\,\sqrt{J_2} \,,{\qquad}
 A_{r2} =  \Bigl(\frac{1}{r - \kp}\Bigr)\Bigl( -\,g\, r + J_2 + \frac{\kp}{2}
 \bigl (r p_r^2 - \frac{J_2}{r}  \bigr)\Bigr) \,,
$$
and 
$$
 B_{\phi 1} =  J_2\cos(2 m\phi)  +  2(k_b-k_a)  \,,{\qquad}
 B_{\phi 2} =   p_\phi\,\sqrt{J_2}\, \sin(2 m\phi) \,.
$$ 

First, we note that the moduli of these two complex functions  are integrals of motion of fourth order in the momenta  given by
\begin{eqnarray*}
  \mid A_r \mid^2  &=&  J_1 J_2 + \bigl(\frac{\kp}{2}\bigr)^2 J_1^2  - \kp\,g J_2 + g^2\cr
  \mid B_\phi \mid^2 &=&  J_2^2 - 4 (k_a+k_b) J_2 +  4 (k_a-k_b)^2 
\end{eqnarray*}

The time-derivative [Poisson bracket with $\wt{H_c}_m(\kp)$] of the function $A_{r1}$ is proportional to $A_{r2}$ and the time-derivative of the $A_{r2}$ is proportional to $A_{r1}$ but with the opposite sign 
$$
 \bigl\{ A_{r1}\,, \wt{H_c}_m(\kp)\bigr\}  =  -\,\lambda_{\kp}\,A_{r2} \,,{\qquad}
 \bigl\{ A_{r2}\,, \wt{H_c}_m(\kp)\bigr\}  =     \lambda_{\kp}\,A_{r1}  \,,
$$
and this property is also true for the angular functions 
$$
 \bigl\{ B_{\phi 1}\,, \wt{H_c}_m(\kp)\bigr\}  = -\,2\,m\,\lambda_{\kp}\,B_{\phi 2} \,,{\qquad}
 \bigl\{ B_{\phi 2}\,, \wt{H_c}_m(\kp)\bigr\}  =    2\,m\,\lambda_{\kp}\,B_{\phi 1} \,,  
$$
where the common factor ${\lambda_\kp}$ takes the value 
$$
 {\lambda_\k} = \frac{1}{( r - \kp)\,r}\,\sqrt{J_2}  \,,{\quad}
 {\lambda_0} = \frac{1}{r^2}\,\sqrt{J_2}   \,. 
$$
Therefore, the time-evolution of the complex functions $A_r$ and $B_\phi$ is given by
$$
 \frac{d}{d t}\,A_r  =    i\, {\lambda_\kp} A_r  \,,{\quad}
 \frac{d}{d t}\,B_\phi  =   i\, 2\,m\,{\lambda_\kp} B_\phi   \,.  
$$
Thus, if we denote by $K_m$ the complex function $K_m=A_r^{(2m)} \,B_\phi^{*} $, then we have
\begin{eqnarray*}
  \frac{d}{dt}\,K_m &=&  \frac{d}{dt}\,\Bigl( A_r^{(2m)} \,B_\phi^{*} \Bigr) 
  = 2 m A_r^{(2m-1)} \,\dot{A_r}\,B_\phi^{*}   
  +   A_r^{(2m)} \,\dot{B_\phi}^{*}    \cr
  &=& A_r^{(2m-1)}\,\Bigl( 2\, m\, ( i\,   {\lambda_\kp}\,A_r)\,B_\phi^{*}   +   A_r\,(-\,  i\, 2\,m\,{\lambda_\kp}\,B_\phi^{*}) \,\Bigr) =  0  \,.
\end{eqnarray*}

We summarize this result in the following proposition. 

\begin{proposicion}    \label{prop6}  
The $\kp$-dependent Hamiltonian $ \wt{H_c}_m(\kp)$ 
\begin{equation}
 \wt{H_c}_m(\kp)  =   \frac{1}{2}\, \Bigl(\frac{r}{r - \kp}\Bigr)   \,\Bigl(\, p_r^2 +  \frac{p_{\phi}^2}{r^2}\,\Bigr)   -\,\frac{g}{ r - \kp} + \frac{F_m(\phi)}{( r - \kp)\,r}   \
\label{HcmF}
\end{equation}
$$
  F_m(\phi) =  \frac{k_a}{\cos^2(m\phi)}  +  \frac{k_b}{\sin^2(m\phi)} \label{Fm}
$$
representing a generalization of the Hamiltonian  $ \wt{H_c}(\kp)$, and also a $\kp$-deformation of the Euclidean PW Hamiltonian $H_{pw}$,  is superintegrable with two quadratic constants of motion and a third constant of motion of higher order. 

\begin{itemize}
\item[(i)]  $\wt{H_c}_m(\kp)$  is separable in polar coordinates  $(r,\phi)$ and it possesses therefore two quadratic constants of motion associated to the Liouville integrability 
\begin{eqnarray*}  
 J_1(\k) &=& \Bigl(\frac{r}{r - \kp}\Bigr)   \,\Bigl(\, p_r^2 +  \frac{p_{\phi}^2}{r^2}\,\Bigr)  -\,\frac{2g}{ r - \kp}+ \frac{2 F_m(\phi)}{( r - \kp)\,r}     \,, \cr 
 J_2 &=& p_\phi^2 + 2 F_m(\phi)   \,. 
\end{eqnarray*}

\item[(ii)] $\wt{H_c}_m(\kp)$ admits a complex constant of motion $K_m$ defined as
$$
  K_m = A_r^{2m} \,B_\phi^{*}    \,. 
$$
\end{itemize}
\end{proposicion}

\section{Final comments }

We observed in the introduction that although the number of superintegrable systems can be considered as rather limited, they are not  however isolated systems but, on the contrary, they frequently appear as grouped in families. 
Now we have proved the existence of four families of Hamiltonians  $\wt{H_r}(\kp)$, $r=a,b,c,d$, associated to previously known super-separable Hamiltonians $H_r$.  
The important point is that the multipler $\lambda$ (that is a function defined in the configuration space) is a continous function of a parameter $\kp$ so that the superintegrability is preserved for all the values of $\kp$ and the integrals of motion (and therefore the associated symmetries) depend in a smooth way of the parameter. The fact that this continous deformations also lead to generalizations of the TTW and the PW systems is certainly a very remarkable property. 

We conclude with the following two comments. 
First, the $\kp$-dependent constants of motion are consequence of the existence of  $\kp$-dependent symmetries; so it would be convenient to study the properties of these symmetries from a geometric approach  (that is, symplectic formalism and Lie algebra of vector fields). 
Second,  it is also convenient to study the quantum versions of these these systems.  It is clear that these Hamiltonians are systems with a position dependent mass (PDM) and therefore the quantization of these systems is not an easy matter. 
These two points are interesting questions deserving to be studied.

\section*{\bf Figures}

\begin{figure}
\centerline{
\includegraphics{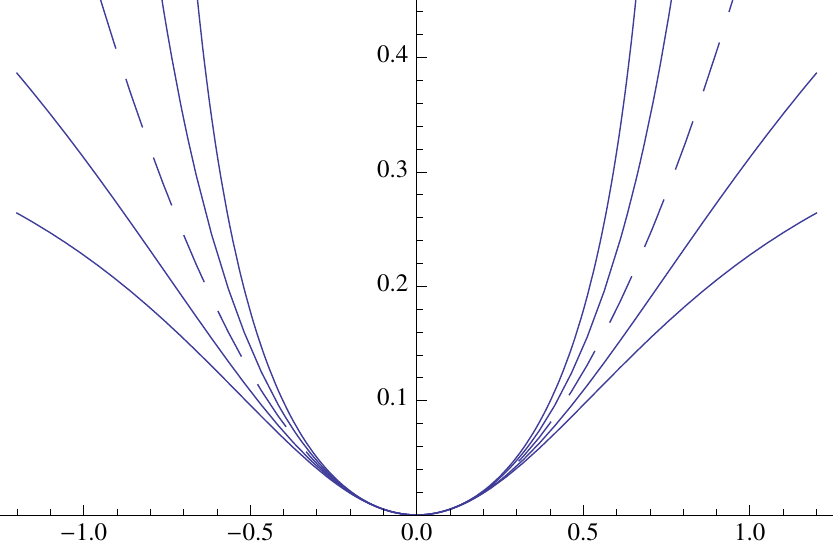}  } 

\caption{Plot of the Potential $V = (1/2)\alpha^2(x^2/(1- \kp\, x^2))$ as a function of $x$,  for $\alpha=1$ and   $\kappa=0$ (dash line), $\kappa>0$ (inner upper curves), and $\kappa<0$ (lower curves in the outside).    }
\label{Fig1}

\end{figure}

\section*{Acknowledgments}

We wish to thank M. Santander for helpful discussions on the theory of super-integrable systems.
This work was supported by the research projects MTM--2012--33575 (MICINN, Madrid)  and DGA-E24/1 (DGA, Zaragoza).

{\small
 }


\begin{thebibliography}{99}
\bibitem{FrMaSmUW65}  T.I. Fris, V. Mandrosov, Y.A. Smorodinsky, M. Uhlir,
and P. Winternitz,
{\rm ``On higher symmetries in quantum mechanics"},
 Phys. Lett.  {\bf 16},  354--356  (1965).
 
\bibitem{Ev90Pra}  N.W. Evans,  
{\rm ``Superintegrability in classical mechanics"},  
Phys. Rev.  A  {\bf  41},  no. 10,  5666--5676  (1990).
\bibitem{GrPoSi95a}  C. Grosche,  G.S. Pogosyan,  and  A.N. Sissakian, 
{\rm ``Path integral discussion for Smoro\-dinsky--Winternitz potentials. 
I two-- and three-- dimensional Euclidean spaces"}, 
Fortschr. Phys.  {\bf  43},  no. 6,  453--521  (1995).
\bibitem{KaWiMiPo99}  E.G. Kalnins, G.C. Williams, W. Miller, and G.S. Pogosyan,  
{\rm ``Superintegrability in the three--dimensional Euclidean space"}, 
J. Math. Phys.  {\bf  40},  no. 2,  708--725  (1999).



\bibitem{Ra97Jmp}  M.F. Ra\~nada,  
{\rm ``Superintegrable $n=2$ systems, quadratic constants and potentials of Drach"},  
J. Math. Phys.  {\bf  38},  no. 8,  4165--4178  (1997).
\bibitem{Tsi00Jpa}  A.V. Tsiganov,
{\rm ``The Drach superintegrable systems"},
J. Phys. A  {\bf  33}, no. 41, 7407--7422  (2000).
\bibitem{RaSa01PLa}  M.F. Ra\~nada and M. Santander, 
{\rm ``Complex euclidean super-integrable potentials, potentials of Drach,  and potential of Holt"},
Phys. Lett. A   {\bf  278},  271--279  (2001). 
\bibitem{Camp14}  R. Campoamor-Stursberg, 
{\rm ``Superposition of super-integrable pseudo-Euclidean  potentials in N = 2 with a fundamental constant of motion of arbitrary order in the momenta"}, 
J. Math. Phys.  {\bf  55},  042904  (2014).



\bibitem{GrPoSi95b}  C. Grosche, G.S. Pogosyan, and A.N. Sissakian, 
{\rm ``Path integral discussion for Smoro\-dinsky--Winternitz potentials. 
II two-- and three-- dimensional sphere"}, 
Fortschr. Phys.  {\bf  43},  no. 6,  523--563  (1995).
\bibitem{RaSa99}  M.F. Ra\~nada and M. Santander,    
{\rm ``Superintegrable systems on the two-dimensional sphere $S^2$ and the 
hyperbolic plane $H^2$"}, 
J. Math. Phys.  {\bf  40},  no. 10,  5026--5057  (1999).
\bibitem{KaKrPoMi01}  E.G. Kalnins, J.M. Kress, G.S. Pogosyan, and W. Miller,  
{\rm ``Completeness of superintegrability in two-dimensional constant-curvature spaces"}, 
J. Phys. A  {\bf  34}, no. 22,  4705--4720  (2001).
\bibitem{KaKrWint02}  E.G. Kalnins, J.M. Kress, and P. Winternitz,  
{\rm ``Superintegrability in a two-dimensional space of nonconstant curvature"}, 
J. Math. Phys. {\bf  43},  no. 2,  970--983  (2002).
\bibitem{BaHeSantS03}  A. Ballesteros,  F.J. Herranz, M. Santander, and T. Sanz-Gil,
{\rm ``Maximal superintegrability on $N$-dimensional curved spaces"},  
J. Phys. A  {\bf  36},  no. 7,   L93--L99  (2003).  
\bibitem{BaHeMu13}  A. Ballesteros,  F.J. Herranz, and Musso F.,  
{\rm ``The anisotropic oscillator on the 2D sphere and the hyperbolic plane"},  
Nonlinearity   {\bf  26},  no. 4,  971--990  (2013).  
\bibitem{GonKas14AnnPhys}  C. Gonera  and  M. Kaszubska,   
{\rm ``Superintegrable systems on spaces of constant curvature"},  
Ann. Physics  {\bf  364},  91--102  (2014). 



\bibitem{MiPWJPa13}  W. Miller,  S. Post,  and  P. Winternitz,  
{\rm ``Classical and quantum superintegrability with applications"},
J. Phys. A: Math. Theor.  {\bf  46},  423001  (2013).     


\bibitem{Woj83}  S.  Wojciechowski, 
{\rm ``Superintegrability of the Caloger-Moser system"}, 
Phys. Lett. A  {\bf  95},  279--281  (1983).
\bibitem{Gon98}  C.  Gonera,  
{\rm ``On the superintegrability of Calogero--Moser--Sutherland model"}, 
J. Phys. A  {\bf 31},  no. 19,   4465--4472  (1998).
\bibitem{Ra99Jmp}  M.F.  Ra\~nada, 
{\rm ``Superintegrability of the Calogero--Moser system: constants of motion, master symmetries, and time-dependent symmetries"}, 
J. Math. Phys.  {\bf  40},  no. 1,  236--247  (1999).


\bibitem{EvVe08b}  N.W. Evans and P.E. Verrier,
{\rm ``Superintegrability of the caged anisotropic oscillator"},
J. Math. Phys.  {\bf  49},  092902  (2008).
\bibitem{RodTW08}  M.A. Rodr\'{\i}guez,  P. Tempesta,  and  P.  Winternitz,
{\rm ``Reduction of superintegrable systems: The anisotropic harmonic oscillator"},
Phys. Rev. E  {\bf 78},  046608  (2008).
\bibitem{RaRoSa10}  M.F. Ra\~nada,  M.A. Rodr\'{\i}guez, and  M. Santander,
{\rm ``A new proof of the higher-order superintegrability of  a noncentral oscillator with inversely quadratic nonlinearities"},
J. Math. Phys.  {\bf  51},  042901  (2010).


\bibitem{TTW09}  F. Tremblay,  A.V. Turbiner, and  P. Winternitz,  
{\rm ``An infinite family of solvable and integrable quantum systems on a plane"},
J. Phys. A: Math. Theor.  {\bf 42},  242001  (2009).
\bibitem{TTW10}  F. Tremblay,  A.V. Turbiner, and  P. Winternitz,  
{\rm ``Periodic orbits for an infinite family of classical super\-integrable systems"},
J. Phys. A: Math. Theor.  {\bf 43},  015202  (2010).


\bibitem{Qu10a}  C.  Quesne,   
 {\rm ``Superintegrability of the Tremblay-Turbiner-Winternitz quantum Hamiltonians on a plane for odd k"},
J. Phys. A: Math. Theor.  {\bf 43},  082001  (2010).
\bibitem{Qu10b}  C.  Quesne,   
{\rm ``N=2 supersymmetric extension of the Tremblay-Turbiner-Winternitz Hamiltonians on a plane"},
J. Phys. A: Math. Theor.  {\bf 43},  305202  (2010).
\bibitem{KaKrMi10}  E.G. Kalnins,  J.M.  Kress,  and  W.  Miller,    
{\rm ``Superintegrability and higher order constants for quantum systems"},  
J. Phys. A: Math. Theor.  {\bf 43},  265205  (2010).
\bibitem{ChDgR11}  A.J.  Maciejewski, M.  Przybylska, and  H.  Yoshida,  
{\rm ``Necessary conditions for super-integrability of a certain family of potentials in constant curvature spaces"}, 
J. Phys. A: Math. Theor.  {\bf  43},  382001  (2010).  
\bibitem{CalCedOlm12}  J.A. Calzada,  E. Celeghini,  M.A. del Olmo,  and M.A. Velasco,  
{\rm ``Algebraic aspects of TTW Hamiltonian system"},
J. Phys. Conf. Series  {\bf  343},  012029  (2012). 
\bibitem{Ra12JPaTTW}  M.F. Ra\~nada,   
 {\rm ``A new approach to the higher-order superintegrability  of the Tremblay-Turbiner-Winternitz system"},
J. Phys. A: Math. Theor.  {\bf 45},  465203  (2012). 
\bibitem{LevPW12JPa}  D. Levesque,  S. Post,  and  P. Winternitz,   
{\rm ``Infinite families of superintegrable systems separable in subgroup coordinates"},  
J. Phys. A: Math. Theor.  {\bf 45},  465204  (2012). 
\bibitem{Gon12PLa}  C. Gonera,  
 {\rm ``On superintegrability of TTW model"},
Phys. Lett. A  {\bf  376},  2341--2343  (2012).  
\bibitem{Hakob12PLa}  T. Hakobyan,  O. Lechtenfeld,  A. Nersessian,  A.  Saghatelian, and V. Yeghikyan,   
{\rm ``Integrable generalizations of oscillator and Coulomb systems via action-angle variables"}, 
Phys. Lett. A  {\bf 376},  679--686  (2012).  

\bibitem{PostTsuVin12}  S. Post,  S. Tsujimoto, and  L. Vinet,   
{\rm ``Families of superintegrable Hamiltonians constructed from exceptional polynomials"}, 
J. Phys. A: Math. Theor.  {\bf 45},  405202  (2012). 

\bibitem{CeKuNedO13}  E. Celeghini, S. Kuru,  J. Negro,  and M.A. del Olmo,  
{\rm ``A unified approach to quantum and classical TTW systems based on factorizations"},  
Ann. Physics {\bf  332},  27--37  (2013).  


\bibitem{CalzadKN14}   J.A.  Calzada, S. Kuru, and J. Negro,   
{\rm ``Superintegrable Lissajous systems on the sphere"},  
Eur. Phys. J. Plus   {\bf  129},   164  (2014). 
\bibitem{Ra14JPaTTWk}  M.F. Ra\~nada,   
 {\rm ``The Tremblay-Turbiner-Winternitz system on spherical and hyperbolic spaces: superintegrability, curvature-dependent formalism and complex factorization"},  
J. Phys. A: Math. Theor.  {\bf  47},   165203  (2014). 


\bibitem{PostWint10}  S.  Post and P.  Winternitz,   
{\rm ``An infinite family of superintegrable deformations of the Coulomb potential"},
J. Phys. A: Math. Theor.  {\bf 43},  222001  (2010). 
\bibitem{Ra13JPaPW}  M.F. Ra\~nada,    
{\rm ``Higher order superintegrability of separable potentials with a new approach to the Post-Winternitz system"}, 
J. Phys. A: Math. Theor.  {\bf 46},  125206  (2013).

 

\bibitem{BurgosPhysd08}  A. Ballesteros,  A. Enciso, F.J. Herranz,  and O. Ragnisco, 
{\rm ``A maximally superintegrable system on an n-dimensional space of nonconstant curvature"},
Phys. D {\bf  237}, no. 4, 505--509  (2008).

\bibitem{BurgosAnnPh11}  A. Ballesteros,  A. Enciso, F.J. Herranz,  O. Ragnisco,  and  D. Riglioni, 
{\rm ``Quantum mechanics on spaces of nonconstant curvature: the oscillator problem and superintegrability"},
 Ann. Physics  {\bf  326},  no. 8,  2053--2073  (2011).

\bibitem{BurgosIntJ11}  A. Ballesteros,  A. Enciso, F.J. Herranz,  O. Ragnisco,  and  D. Riglioni,
{\rm ``On two superintegrable nonlinear oscillators in N dimensions"}, 
Internat. J. Theoret. Phys. {\bf  50},  no. 7,  2268--2277  (2011). 

\bibitem{BurgosSigma11}  A. Ballesteros,  A. Enciso,  F.J. Herranz,  O. Ragnisco, and  D. Riglioni, 
{\rm ``Superintegrable oscillator and Kepler systems on spaces of nonconstant curvature via the St\"ackel transform"},  
SIGMA (Symmetry Integrability Geom. Methods Appl.)  {\bf  7}, paper 048  (2011). 


\bibitem{MatLak74} P.M. Mathews and M. Lakshmanan,  
{\rm ``On a unique nonlinear  oscillator"}, 
 Quart. Appl. Math. {\bf  32},  215--218 (1974). 
\bibitem{ChSentKak05}  Chandrasekar V.K., Senthilvelan M., and  Lakshmanan M., 
``Unusual Lienard-type nonlinear oscillator",
Phys. Rev. E  {\bf  72}, no.  6,  066203  (2005).
\bibitem{BruGanSen11}  M.S. Bruzon, M.L. Gandarias, and M. Senthilvelan, 
{\rm ``On the  nonlocal symmetries of certain nonlinear oscillators and their 
general solution"},
Phys. Lett. A  {\bf 375},  2985--2987 (2011). 
\bibitem{CadLRan15}  J.C. Cari\~nena,  J. de Lucas,  and  M.F. Ra\~nada,
{\rm ``Jacobi multipliers, non-local symmetries and  nonlinear oscillators"}
  (to be published). 


\bibitem{CaRaSS04}  J.F. Cari\~nena, M.F. Ra\~nada, M. Santander, and M. Senthilvelan, 
{\rm ``A non-linear Oscillator with quasi-Harmonic behaviour:
two- and $n$-dimensional oscillators"},
Nonlinearity {\bf 17},  no. 5,  1941--1963 (2004).
\bibitem{CaRaSaAnPh07a}  J.F. Cari\~nena,  M.F. Ra\~nada, and M. Santander, 
{\rm ``A quantum exactly solvable nonlinear oscillator with quasi-harmonic behaviour"},
Ann. of  Physics  {\bf 322},  no. 2,  434--459 (2007).

\bibitem{CaRaSaAnPh07b}  J.F. Cari\~nena,  M.F. Ra\~nada, and M. Santander, 
{\rm ``The quantum harmonic oscillator on the sphere and the hyperbolic plane"},
Ann. of  Physics  {\bf 322},  no. 10,  2249--2278 (2007).
\bibitem{CaRaSaIntJ11}  J.F. Cari\~nena,  M.F. Ra\~nada, and M. Santander, 
{\rm ``The harmonic oscillator on three-dimensional spherical and  hyperbolic spaces:  Curvature dependent formalism and quantization"}, 
Int. J. Theoretical Physics  {\bf  50},  2170--2178  (2011).   
\bibitem{CaRaSaJPh12}  J.F. Cari\~nena,  M.F. Ra\~nada, and M. Santander, 
{\rm ``Curvature-dependent formalism, Schrodinger equation and energy levels for the harmonic oscillator  three-dimensional spherical and  hyperbolic spaces"}, 
J. Phys. A: Math. Theor.  {\bf  45},  265303  (2012). 


\bibitem{CaRaSaJmp05}  J.F. Cari\~nena,  M.F. Ra\~nada, and M. Santander, 
{\rm ``Central potentials on spaces of constant curvature:
the Kepler problem on the two- dimensional sphere $S^2$ and
the hyperbolic plane $H^2$"}, 
J. Math. Phys. {\bf 46},  no. 5,  052702   (2005).
\bibitem{CaRaSaJpa07}  J.F. Cari\~nena,  M.F. Ra\~nada, and M. Santander, 
{\rm ``Superintegrability on curved spaces, orbits and momentum hodographs:
revisiting a classical result by Hamilton"}, 
J. Phys. A: Math. Theor.  {\bf  40}, no. 45, 13645--13666  (2007).


\bibitem{KaMiPost10}  E.G. Kalnins,  W. Miller,  and  S. Post,
{\rm ``Coupling constant metamorphosis and $N$th-order symmetries in classical and quantum mechanics"},
J. Phys. A: Math. Theor.  {\bf 43},  035202  (2010). 
 

\end{thebibliography}
\end{document}